\documentclass[11pt]{article}

\usepackage{amsmath,amsfonts,amssymb,amsthm,latexsym}
\usepackage[mathscr]{eucal}
\usepackage{graphicx} 
\usepackage{textcomp}
\newtheorem{definition}{Definition}

\begin{document}

\centerline{\Large \bf  Polynomial Approximations  }

\centerline{\Large \bf of Electronic Wave Functions}

\bigbreak \bigbreak

\centerline {A. I. Panin}

\centerline{ \footnotesize Chemistry Department, St.-Petersburg State University,}

\centerline {\footnotesize  St.-Petersburg 198504, Russia }

\centerline {\footnotesize  e-mail:  andrej.panin@gmail.com}

\bigbreak

\newpage

\bigbreak

{\bf ABSTRACT: }{ This work completes the construction of purely algebraic version of the theory of non-linear quantum chemistry methods. It is shown that at the heart of these methods there lie certain algebras  close in their  definition to the well-known Clifford algebra  but quite different in their properties. The most important for quantum chemistry property of these algebras is the following : for a fixed number of electrons the corresponding sector of the Fock space becomes a commutative algebra and its  ideals are determined by the order of  excitations  from the Hartree-Fock reference state. Quotients of this algebra can also be endowed with commutative algebra structures and quotient Schr{\"o}dinger equations are exactly the couple cluster type  equations. Possible computer implementation of multiplication in the aforementioned algebras is described. Quality of different polynomial approximations of configuration interaction wave functions is illustrated with concrete examples. Embedding of algebras of infinitely separated subsystems in algebra of the united system is discussed.   }

\bigbreak {\bf Key words: }{\small Fock space; commutative algebras;  non-linear wave function based methods   }

\bigbreak

\hrule
\bigbreak
{\Large \bf Introduction}  
\bigbreak

At present stage of development the non-linear methods of quantum chemistry take on spectial significance in  investigation of molecular systems when electronic correlation effects should be accounted on rigorous and detailed level. Among such methods non-variational version of Coupled Cluster (CC) method is most popular now. Its theoretic backgrounds were formulated in the seventies of the last century \cite{Cizek-1, Cizek-2} and gained development in further numerous publications (see \cite{Bartlett-1}-\cite{Crawford} and references therein).

There were attempts of purely algebraic approach to non-linear theories of quantum chemistry (see, e.g., \cite{Paldus-2,  Paldus-3}) but with moderate success. And only after pubication of papers \cite{Panin-1,Panin-2} it became clear that at the heart of non-linear methods of CC type there lies  a very special but purely algebraic structure close in its definition to the well-known Clifford  algebra (see, e. g., \cite{Lang, Kost}) but essentially different in its properties. And it became possible to give a uniform description of non-linear methods of quantum chemistry on modern algebraic level and compare their quality. 

In present paper we give a final reasonably rigorous report on algebraic backgrounds of practically all existing non-linear methods of quantum chemistry and on concrete examples compare quality of different polynomial approximations of many electron wave functions.

\newpage
{\Large \bf \noindent Formal Fock Space} \bigbreak \bigbreak

\bigbreak \bigbreak

In quantum theory  notion of space of occupation numbers  is widely used and became habitual both to physicists and to quantum chemists (see, e.g., \cite{March, Raimes}). Recall that if the number of molecular spin orbitals (MSO) is finite and if they are labelled by  indices from the index set $N=\{1,2,\ldots,n\}$ then the corresponding  space of occupation numbers is generated by bit vectors $(n_1,n_2,\ldots )$ of length $n$ with $n_i=0(1)$. These bit vectors constitute a basis of $2^n$ - dimensional vector space and their interpretation is very simple: each bit vector corresponds to {\it  normalized}  Slater determinant built on MSOs having non-zero occupancies in the bit vector under consideration. Support of bit vector is defined as a subset of the index set $N$, involving indices (positions) where occupation number is equal to 1. It is clear that each bit vector is determined by its support and \textsl{vice versa}. It is possible therefore to replace bit vectors by their supports and consider  linear combinations of all subsets of the index set $N$.      
\begin{definition} Vector space of formal linear combinations of all subsets of the index set $N$ with coefficients from the ground field $\mathbb K$ will be called the formal Fock space over $\mathbb K$ associated with the index set $N$ and will be denoted by the symbol ${\mathfrak F}_N$.  
\label{formal}
\end{definition}
Of course, in this work  fields of real ($\mathbb K=\mathbb R$) and complex ($\mathbb K=\mathbb C$) numbers  will be of actual interest.
 
For each subset $R\subset N$ symbol $e_R$ will stand for a basis vector of the formal Fock space. Of course, $e_R$ is just a notation for a subset of the index set $N$, playing, as used here,  a role of a basis vector. 

Subsets with the same number of indices generate subspace (sector) of the formal Fock space and 
\begin{equation}
{\mathfrak F}_N=\bigoplus_{p=0}^n{\mathfrak F}_N^p
\end{equation}   
where
\begin{eqnarray}
{\mathfrak F}_N^p=\bigoplus_{R\subset N}^{(p)}{\mathbb K}e_R
\end{eqnarray}
With each index from the MSO index set  creation and annihilation operators can be associated and formal Hamiltonians can be defined in terms of these creation-annihilation operators and 1 - and 2 - electron integrals as parameters.

Definition of formal Fock space immediately suggests that one can try to use standard set-theoretic operations (union, intersection, difference, symmetric difference)  to introduce in this space new  laws of composition in addition to the already existing (addition and multiplication by scalars from $\mathbb K$). It should be emphasized that such an approach is widely used in modern mathematics. Namely, first a set  with some algebraic operations on it is taken,  then vector space of formal linear combinations  of elements of this set is constructed and initial algebraic operations  are  continued by linearity to the space of formal linear combinations.  The standard  example is a formal linear hull ${\mathbb K}[G]$ of multiplicative group $G$. Putting  
\begin{equation}
(\sum\limits_{g\in G}a_ge_g)(\sum\limits_{h\in G}b_he_h)=\sum\limits_{g,h\in G}a_gb_he_{gh}
\end{equation}
we come to the well-known and very important object called group algebra of (finite) group $G$ over $\mathbb K$ (see, e.g., \cite{Serre}).

We start with a law of composition (multiplication) on ${\mathfrak F}_N$ which is well-known for about 120 years and was first introduced by Clifford \cite{Clifford}. Its modern interpretation  may be found in \cite{Lang, Kost}. This multiplication 
is a combination of  set-theoretic union of  MSO indices and  contraction of a certain part of these indices with the aid of some fixed symmetric bilinear form $g$ on  '1-electron` space ${\mathfrak F}^1_N$. It is convenient to assume that  basis constituted by the vectors $e_i=e_{\{i\}}$ is orthogonal with respect this form. Then the aforementioned law of composition is defined as \cite{Lang, Kost}
\begin{equation}
e_R.e_S=\prod\limits_{\genfrac{}{}{0pt}{}{r\in R}{s\in S}
}(r,s)\prod\limits_{i\in R\cap S}g(e_i,e_i)e_{R\Delta S}
\label{clif}
\end{equation}
where $R\Delta S=R\cup S-R\cap S$ and 
\begin{equation}
(r,s)=\begin{cases}
\ \ 1 &\text{if}\quad r\le s\cr
-1 &\text{if}\quad r>s
\end{cases}
\end{equation}
and where it is assumed that 'empty` products are equal to 1. In terms of introduced in our previous publications set-theoretic operation   $\Delta_K=\Delta_{k\in K}[1,k]$ (see, e.g., \cite {Panin-1}) the sign prefactor may be written as
\begin{equation}
\prod\limits_{\genfrac{}{}{0pt}{}{r\in R}{s\in S}
}(r,s)=(-1)^{|(S-R)\cap\Delta_{(R-S)}|}
\end{equation}
This presentation of sign prefactor is very convenient both for analytic and numerical purposes. For example, proof of  associativity in Clifford-type  algebras becomes practically automatic.

Directly from definition it follows that $e_{\text{\O}}$ is the identity element of the multiplication (\ref{clif}).  
It is easy to show that this law of composition  is associative, non-commutative,  and endows the formal Fock space with the Clifford algebra structure. Classic relation 
\begin{equation}
x.y+y.x=2g(x,y)e_{\text{\O}},
\end{equation}  
valid for any elements $x,y\in {\mathfrak F}^1_N$,   is readily obtained from Eq.(\ref{clif}).

If $g$ is zero form then the Clifford algebra is just a Grassmann algebra of '1-electron' space ${\mathfrak F}^1_N$. 

Another type  of multiplications on the formal Fock space was introduced in our previous publications \cite{Panin-1, Panin-2}. Recall briefly most important definitions.

For any fixed subset $R\subset N$ ('reference subset') basis vectors of the formal Fock space may be written  as $e_J^I(R)=e_{R-J\cup I}$ where $J\subset R$, $I\subset N-R$. Law of composition of these basis vectors ('star product') is defined as (see \cite{Panin-1})
\begin{equation}
e_J^I(R)* e_{J'}^{I'}(R)= 
\begin{cases}
\left [\prod\limits_{\genfrac{}{}{0pt}{}{r\in I\cup J}{s\in I'\cup J'}
}(r,s)\right ] e_{J\cup J'}^{I\cup I'}(R) & \text{if $(I\cup J)\cap (I'\cup J')=\text{\O}$},\\
\qquad\quad 0 & \text{if $(I\cup J)\cap (I'\cup J')\ne\text{\O}$} 
\end{cases}
\label{multtab}
\end{equation}

Relations (\ref{multtab}) constitute the multiplication table of an algebra and it is easy to ascertain that this algebra is associative, skew-commutative, and that $e_{\text{\O}}^{\text{\O}}(R)=e_R$ is its identity element. Skew-commutativity is readily follows from the equality
\begin{equation}
e_J^I(R)* e_{J'}^{I'}(R)= 
(-1)^{|J\cup I|\cdot |J'\cup I'|}e_{J'}^{I'}(R)*e_J^I(R)
\label{skew}
\end{equation}  

In the finite-dimensional case we have $2^n$ ways to select reference subset and, consequently, $2^n$ different multiplications on the formal Fock space. Of course, algebras, corresponding to  reference subsets with the same number of electrons are isomorphic. 

The simplest reference subset is $R=\text{\O}$ and the corresponding basis vectors are $e_{\text{\O}}^I(\text{\O})=e_I$. We have the following relations
\begin{equation}
e_{\text{\O}}^I(\text{\O})* e_{\text{\O}}^{I'}(\text{\O})= 
\begin{cases}
\left [\prod\limits_{\genfrac{}{}{0pt}{}{r\in I}{s\in I'}
}(r,s)\right ] e_{\text{\O}}^{I\cup I'}(\text{\O}) & \text{if $I\cap I'=\text{\O}$},\\
\qquad\quad 0 & \text{if $I\cap I'\ne\text{\O}$} 
\end{cases}
\label{multgrass}
\end{equation}  
which constitute the multiplication table of the Grassmann algebra. In this particular case the star product is just the classic Grassmann exterior product. 

If the reference subset $R$ belongs to the $p$-electron sector of the formal Fock space ($|R|=p>0$) and $|I|=|J|=r\le \min\{p,n-p\}$ then, as immediately follows from Eq.(\ref{skew}), the star product endows this sector with the {\it commutative} algebra structure. As will be seen later, this structure is particularly important for quantum chemistry. Unless otherwise stated, from here on only $p$-electron sector ${\mathfrak F}_N^p$ of the formal Fock space will be under consideration. For fixed $p$-element subset $R$ the commutative $p$-electron algebra with the identity $e_R$ will be denoted ${ \rm A}_N^p(R)$. Its basis elements are $e_J^I(R)$ with $|J|=|I|=l$ where $l$ is the so-called {\it excitation order or excitation level} ($l=0,1,\ldots,\min\{p,n-p\}$).

Algebra ${ \rm A}_N^p(R)$ admits a decomposition into a direct sum of subspaces
\begin{equation}
{ \rm A}_N^p(R)=\bigoplus_{l\in \mathbb Z}{\rm V}_N^l(R)
\end{equation}
where
\begin{equation}
{\rm V}_N^l(R)=\bigoplus\limits_{\genfrac{}{}{0pt}{}{J\subset R}{I\subset N-R}}^{(l)}{\mathbb K}e_J^I(R)
\end{equation}
and where ${\rm V}_N^l(R)=\{0\}$ for $l<0$ and $l> \min\{p,n-p\}$. For fixed $1\le l\le \min\{p,n-p\}$ this subspace
can be called `subspace of all excitation of order $l$ from the reference subset $R$'. From the definition of the star product it readily follows that
\begin{equation}
{\rm V}_N^k(R)*{\rm V}_N^l(R)\subset {\rm V}_N^{k+l}(R)
\end{equation}
which means that ${ \rm A}_N^p(R)$ is \textsl{a $\mathbb Z$-graded algebra}. 

Subspace 
\begin{equation}
{\rm I}_N^r(R)=\bigoplus_{l=r+1}^{\min\{p,n-p\}}{\rm V}_N^l(R)
\end{equation}
is an ideal of ${ \rm A}_N^p(R)$ for any $r\ge 0$. This means that any  quotient space  ${ \rm A}_N^p(R)/{\rm I}_N^r(R)$ is also a commutative algebra with the law of composition
\begin{equation}
[\Psi_1*\Psi_2]_r=[\Psi_1]_r*[\Psi_2]_r
\end{equation}
where  $[\Psi]_r=\Psi+{\rm I}_N^r(R)$. 
By an abuse of language star product in the quotient algebra ${ \rm A}_N^p(R)/{\rm I}_N^r(R)$ will be called 'star product modulo $r$`.

It is clear that ${\rm I}_N^0(R)$ is \textsl{a maximal nilpotent ideal} of ${ \rm A}_N^p(R)$. Maximality means that it is not contained in any ideal of the algebra under consideration and nilpotency means that $\left [{\rm I}_N^0(R)\right ]^r={\rm I}_N^0(R)*\ldots *{\rm I}_N^0(R)=\{0\}$ for $r > \min\{p,n-p\}$.

${\rm A}_N^p(R)$ is an abstract model of a typical full Configuration Interaction (CI) space (endowed with the additional multiplication) devoid of any orbital specificity. Vectors from the vector space 
\begin{equation}
{\rm W}_N^r(R)=\bigoplus\limits_{l=1}^r{\rm V}_N^l(R)
\end{equation}
will be called amplitudes. \textsl{Note that in contrast to the standard CC type methods our amplitudes are just wave functions from the same $p$ - electron sector of the Fock space.}

If ${\cal F}^1$ is a concrete 1-electron space and $\psi=(\psi_1,\psi_2,\ldots,\psi_n)$ is some its orthonormal $n$-frame of MSOs, then  the {\it substitution mapping}
\begin{equation}
s_{\psi}:e_R\mapsto \psi_{i_1}\wedge\psi_{i_2}\wedge\ldots\wedge\psi_{i_p}, 
\end{equation}
where $R=\{i_1,i_2\ldots,i_p\}$, $i_1<i_2<\ldots <i_p$ establishes (basis-dependent)  isomorphism between the formal Fock space ${\mathfrak F}_N$  and  concrete Fock space 
\begin{equation}
{\cal F} = \bigoplus_{p=0}^n\bigwedge^p {\cal F}^1 
\end{equation}
Substitution mapping can also be used to transfer the algebra structures defined by Eq.(\ref{multtab}) from the formal Fock space to any its concrete realization.  The inverse mapping $s_{\psi}^{-1}$ (also basis dependent) strips vectors of concrete Fock space of their orbital specificity. 

\textsl{It pertinent to emphasize that for any concrete realization of the Fock space star product of arbitrary $p$ - electron wave functions is (antisymmetric) $p$ - electron wave function.}
  
The formal Fock space can be endowed with Euclidean ($\mathbb K=\mathbb R$) or Hermitian ($\mathbb K=\mathbb C$) structure in a simple and natural way:
\begin{equation}
\langle e_R|e_{R'}\rangle = 
\begin{cases}
1 & \text{if $R=R'$}\\
0 & \text{if $R\ne R'$} 
\end{cases}
\label{innerp}
\end{equation}     
If a concrete 1-electron Fock space is a vector space with the inner product and $\psi$ is some its orthonormal (with respect to this inner product)  $n$-frame then the substitution mapping is obviously isometric. 

The group of orthogonal (or  unitary) transformations of 1-electron Fock space ${\cal F}^1$ will be denoted   ${\cal G}({\cal F}^1)$. If $\varpi_R$ is $p$-plane ($p$-dimensional subspace) of 1-electron Fock space spanned by MSOs $\{\psi_i\}, i\in R$ and $\varpi_R^{\perp}$ is its orthogonal complement then for any block-diagonal transformation $U\in {\cal G}(\varpi_R)\times {\cal G}(\varpi_R^{\perp})$
\begin{eqnarray}
s_{\psi}({\rm V}_N^l(R))=s_{\psi U}({\rm V}_N^l(R))
\label{invar} 
\end{eqnarray}
where $l=0,1,\ldots,\min\{p,n-p\}$. In the case $l=0$ this equality is equivalent to the well-known property of the determinant states: each determinant state is defined up to arbitrary orthogonal (unitary) transformation of occupied (virtual) MSOs. Therefore, Eq.(\ref{invar}) may be considered as generalization of this simple property of determinants.  It is to be noted that, in general, $s_{\psi}(\tau)\ne s_{\psi'}(\tau)$ for \textsl{concrete} non-zero vector $\tau\in {\rm V}_N^l(R)$.        

\bigbreak \bigbreak
{\Large \bf \noindent Polynomial Functions With  Algebra ${ \rm A}_N^p(R)$ As Their Range}
\bigbreak \bigbreak

\bigbreak \bigbreak

Many  methods of quantum chemistry may be uniformly described in terms of polynomial functions $P_a:{\rm W}_N^r(R)\to {\rm A}_N^p(R)$ of the form 
\begin{equation}
 P_a(\tau)=\sum\limits_{\mu=0}^{\min\{p,n-p\}}a_{\mu}{\tau}^{\mu}
\label{polfunc} 
\end{equation}
where $a_i\in \mathbb K$ and $\tau^k=\underbrace{\tau *\cdots *\tau }_{k\  times}$. Since ${\rm W}_N^r(R)$ is a subspace of the nilpotent ideal ${\rm I}_N^0(R)$, all   polynomial functions of the type of Eq.(\ref{polfunc}) have degrees not greater than $\min\{p,n-p\}$.

Using polynomial functions it is easy to characterize invertible elements of algebra ${\rm A}_N^p(R)$.

{\bf Propsition.}
\textsl{Wave function $\Psi$ is invertible element of algebra ${\rm A}_N^p(R)$ if and only if $a_0=\langle \Psi|e_{\text{\O}}^{\text{\O}}(R)\rangle \ne 0$.}

{\bf Proof.} Let us put $\tau=\Psi-a_0e_{\text{\O}}^{\text{\O}}(R)$. By direct calculation it is easy to ascertain that 
\begin{equation}
\Psi^{-1}=\sum\limits_{\mu=0}^{\min\{p,n-p\}}(-1)^{\mu}\frac{\tau^{\mu}}{a_0^{\mu+1}}
\end{equation}

{\bf Corollary.}
\textsl{Amplitude vectors are not invertible.}

Polynomial functions $P_a(\tau)$ such that (1) $a_0\ne 0$, and (2) $P_a(\tau)$ is injective, are most important for applications in quantum chemistry. The first condition means that for any amplitude vector $\tau$ the state vector $P_a(\tau)$ has non-zero projection on the reference basis vector $e_{\text{\O}}^{\text{\O}}(R)$, and, consequently, is  invertible.  Without loss of generality we can put  $a_0=1$.  The second condition guarantees that the mapping $\tau\to P_a(\tau)$ is \textsl{a parametrization of  the surface} ${\cal S}^r_N(R)=P_a\left ({\rm W}_N^r(R)\right )$ in the algebra ${\rm A}_N^p(R)$ of $p$-electron states. Dimension of this surface is equal to the dimension of the amplitude space ${\rm W}_N^r(R)$. 

{\bf Propsition.} Inverse of polynomial function $P_a(\tau):{\rm W}_N^r(R)\to {\cal S}^r_N(R)$ is also a polynomial function and it exists if and only if $a_1\ne 0$.

{\bf Proof.} Follows directly from the recurrence formulas for the coefficients of the inverse polynomial $P_a^{-1}(a_0e_{\text{\O}}^{\text{\O}}(R)+\tau)=\sum\limits_{\mu=1}(-1)^{\mu-1}b_{\mu}\tau^{\mu}$:
\begin{subequations}
\begin{equation}
b_{\mu}=\frac{(-1)^{\mu}}{a_1^{\mu}}\sum\limits_{\nu=1}^{\mu-1}(-1)^{\nu-1}b_{\nu}
\sum\limits_{i_1+\cdots +i_{\nu}=\mu}a_{i_1}\cdots
a_{i_{\nu}},\end{equation}
\begin{equation}
 b_1=\frac{1}{a_1}.
\end{equation}
\label{inv_coef}
\end{subequations}

Note that invertibility of $P_a(\tau)$  is equivalent to the condition that the polynomial derivative 
\begin{equation}
P'_a(\tau)=\left (\sum\limits_{\mu=0}^{\min\{p,n-p\}}a_{\mu}{\tau}^{\mu}\right )'=\sum\limits_{\mu=1}^{{\min\{p,n-p\}}}\mu a_{\mu}{\tau}^{\mu -1}
\label{polder}
\end{equation}
is non-zero for any amplitude vector $\tau$ as could be expected.

For any reference subset $R$ and any polynomial function  $P_a(\tau)$ with non-zero $a_0$ and $a_1$ a triple $c_R=\left ({\cal S}^r_N(R),[P_a]^{-1},{\rm W}_N^r(R)\right )$ is a chart of a geometric object which can be called \textsl{polynomial manifold}. Note that any point of the surface ${\cal S}^r_N(R)$ is invertible in algebra ${\rm A}_N^p(R)$. If wave function $\Psi$ belongs to the surface ${\cal S}^r_N(R)$ then $\Psi^{-1}$ belongs to the surface which can be called 'surface of inverse elements of ${\cal S}^r_N(R)$` and, by a certain abuse of language, will be called surface inverse to ${\cal S}^r_N(R)$. \textsl{Thus, each polynomial surface  
(manifold) arises in pair with its inverse. } 

It is pertinent to note that for a fixed polynomial parametrization all arising surfaces form a chain
\begin{equation}
{\cal S}^1_N(R)\subset {\cal S}^2_N(R)\subset \ldots \subset {\cal S}^{\min\{p,n-p\}}_N(R)
\end{equation}
and, consequently, any surface ${\cal S}^r_N(R)$ can be treated as a subsurface of ${\cal S}^{r'}_N(R)$
for any $r'>r$. As a result, it is possible to take some reference $\tau_0$ corresponding to a fixed linear combination of determinants and involving certain excitations of \textsl{arbitrary} order from the reference subset $R$. Considering $P_a(\tau_0)$ as the origin, it is possible to construct parametrizations of the type $\tau\mapsto P_a(\tau_0)*P_a(\tau)$ where $\tau$ include all excitations up to preferable maximal order $r$. If $\tau_0=0$, we come to the standard  polynomial parametrization. Such an approach presupposes that $\tau_0$ is sparse to be kept in fast memory. 

Among {\it infinitely many} polynomial functions  at present only three types  are of actual use in quantum chemistry. The first type is certainly the  affine function
\begin{equation}
P_{CI}:\tau\to e_{\text{\O}}^{\text{\O}}(R)+\tau,\qquad P^{-1}_{CI} :e_{\text{\O}}^{\text{\O}}(R)+\tau\to \tau
\end{equation}
which corresponds to the CI approach accounting all excitations up to order $r$ from the reference index set $R$. The corresponding surface ${\cal S}C^r_N(R)$ is just \textsl{the affine plane} $e_{\text{\O}}^{\text{\O}}(R)+{\rm W}^r_N(R)$. The inverse surface is
\begin{equation}
\tau\mapsto\sum\limits_{k=0}^{\min\{p,n-p\}}(-1)^k\tau^k
\end{equation}

The second type is the exponential function
\begin{equation}
\exp:\tau\to \sum\limits_{{\mu}=0}^{{\min\{p,n-p\}}}\frac{{\tau}^{\mu}}{{\mu}!},\quad  \exp^{-1}: e_{\text{\O}}^{\text{\O}}(R)+x\to \sum\limits_{\mu=1}^{{\min\{p,n-p\}}}
(-1)^{\mu-1}\frac{{x}^{\mu}}{\mu}
\label{explog}
\end{equation}
which appears in coupled cluster  methods. Exponential surface  ${\cal S}E^r_N(R)$ coincides with its inverse and is, in fact,\textsl{ a multiplicative Abelian group}. 

And the last, third type is 
\begin{equation}
q_{\alpha}:\tau\mapsto  e_{\text{\O}}^{\text{\O}}(R)+\tau+\alpha\tau^2, \quad  q_{\alpha}^{-1}: e_{\text{\O}}^{\text{\O}}(R)+x\to x-\alpha x^2+\frac{12\alpha^2-1}{6}x^3+\cdots
\end{equation}
corresponding to the quadratic CI (QCI) method \cite{Pople}. With such a parametrization calculation of polynomial value  can be done very fast but to reach reasonable precision in approximation of wave function  excitation level should be sufficiently  high. 

Parametrization closely related to the resolvent mapping in algebra  ${ \rm A}_N^p(R)$ can also be used
\begin{equation}
Q:\tau\to\sum\limits_{i=0}^{{\min\{p,n-p\}}}\tau^i, \qquad  Q^{-1}: e_{\text{\O}}^{\text{\O}}(R)+x\to \sum\limits_{\mu=1}^{{\min\{p,n-p\}}}(-1)^{\mu -1}x^{\mu}
\end{equation}
\textsl{It is easy to see that  surfaces ${\cal S}Q^r_N(R)$ and  ${\cal S}C^r_N(R)$ are mutually inverse.}

If $r=\min\{p,n-p\}$ then {\it any} injective polynomial function  may be used to parametrize \textsl{all} $p$-electron states that have non-zero projection on the reference basis vector $e_{\text{\O}}^{\text{\O}}(R)$. This means that from the variational viewpoint all the aforementioned polynomial parametrizations are equivalent. If $r<\min\{p,n-p\}$ then different polynomial functions parametrize different \textsl{low dimensional surfaces} in the space of all $p$-electron states and correspond, in general, to different variational methods. {\it In this case the choice of appropriate polynomial parametrization is of primary importance}.  

\bigbreak \bigbreak
{\Large \bf \noindent Variational Methods on Polynomial Surfaces}
\bigbreak \bigbreak

\bigbreak \bigbreak

In this section, unless otherwise indicated, it will be  supposed that the reference index set $R$ is fixed, and  symbol $R$ in all expressions will be supressed. 

For arbitrary polynomial function $P_a(\tau)$ the electronic energy expression may be written in two equivalent forms: 
\begin{subequations}
\begin{eqnarray}
\label{energya}
E_a(\tau)&=&\frac{\langle P_a(s_{\psi}(\tau))|H| P_a(s_{\psi}(\tau))\rangle}{\|P_a(\tau)\|^2}\\
E_a(\tau)&=&\frac{\langle P_a(\tau)|H_{\psi}|P_a(\tau)\rangle}{\|P_a(\tau)\|^2}
\label{energyb}
\end{eqnarray}
\label{energy}
\end{subequations}
where $\psi$ is arbitrary $n$-frame of ${\cal F}^1$, $H$ is the standard $p$-electron Hamiltonian, and $H_{\psi}$ is its {\it formal} parametric analogue expressed via {\it formal} creation-annihilation operators
\begin{equation}
H_{\psi}=\sum\limits_{i,j=1}^n\langle\psi_i|h|\psi_j\rangle{\rm a}^{\dagger}_i{\rm a}_j+\frac{1}{2}\sum\limits_{i,j,k,l=1}^n\langle\psi_i\psi_j|\psi_k\psi_l\rangle{\rm a}^{\dagger}_i{\rm a}^{\dagger}_j{\rm a}_l{\rm a}_k
\label{formalH}
\end{equation}
Remind that $s_{\psi}$ is isometry and, consequently, $\|P_a(s_{\psi}(\tau))\|^2=\|P_a(\tau)\|^2$. 

Unless otherwise indicated, the energy expression (\ref{energyb}) will be used. 

To get energy stationary conditions, it is necessary to find out interrelation between star product and derivation. 
For $\mathbb K=\mathbb R$ electronic energy  is analytic function of amplitudes whereas for $\mathbb K=\mathbb C$ it is not. However, realification of the complex amplitude space  improves the situation. Let us put 
\begin{equation}
\tau=\sum\limits_{l=1}^r\sum\limits_{\genfrac{}{}{0pt}{}{J\subset R}{I\subset N-R}}^{(l)}(x_J^I+{\rm\bf i}y_J^I)e_J^I=x+{\rm\bf i}y
\end{equation}
Then 
\begin{equation}
\frac{\partial}{\partial x_J^I}P_a(\tau)=P'_a(\tau)*e_J^I,\qquad \frac{\partial}{\partial y_J^I}P_a(\tau)={\rm\bf i}P'_a(\tau)*e_J^I
\end{equation}
and
\begin{subequations}
\begin{eqnarray}
\frac{\partial E_a}{\partial x_J^I} (\tau)= \frac{2}{\|P_a(\tau)\|^2}{\rm Re}\langle P'_a(\tau)*e_J^I|H_{\psi}-E_a(\tau)I|P_a(\tau)\rangle\\
\frac{\partial E_a}{\partial y_J^I} (\tau)= \frac{2}{\|P_a(\tau)\|^2}{\rm Im}\langle P'_a(\tau)*e_J^I|H_{\psi}-E_a(\tau)I|P_a(\tau)\rangle
\end{eqnarray}
\end{subequations}   
where the expression for polynomial derivative $P'_a(\tau)$ is given by Eq.(\ref{polder}). 
 
Now we can write down stationary conditions for electronic energy as a function on the surface ${\cal S}^r_N$:
\begin{equation}
\langle P'_a(\tau)*e_J^I|H_{\psi}-E_a(\tau){\hat I}|P_a(\tau)\rangle =0
\label{statcond}
\end{equation}
where $J\subset R$, $I\subset N-R$, $|J|=|I|=1,2,\ldots,r$.

Wave functions $P'_a(\tau)*e_J^I$ and ${\rm\bf i}P'_a(\tau)*e_J^I$ are linearly independent forming a basis of the realified tangent space ${\sf T}_{P_a(\tau)}{\cal S}^r_N$ to the surface ${\cal S}^r_N$ an the point $P_a(\tau)$.

In the particular case $P_a(\tau)=e_{\text{\O}}^{\text{\O}}+\tau$   system (\ref{statcond})  is equivalent to the eigenvalue problem for the projection of the electronic Hamiltonian on the subspace spanned by the basis vectors $e_J^I$ with $0\le|I|=|J|\le r$ (CI method accounting all excitations up to order $r$ from the reference determinant). In general case  the system (\ref{statcond}) can be reduced  to eigenvalue problems for electronic Hamiltonian projection on $\tau$ - dependent subspace spanned by vector $P_a(\tau)$ and tangent vectors $P'_a(\tau)*e_J^I$. Since these vectors are not orthogonal, non-trivial Gram (overlap) matrix arises and  the aforementioned eigenvalue problem is actually a generalized one. 

At this stage we can see advantages of the exponential parametrization. First, calculation of the polynomial derivative is not required because $\exp'(\tau)*e_J^I=\exp(\tau)*e_J^I$. Second, algorithm for $\tau$ update is very simple and efficient and can be described as follows.

(i) For a fixed amplitude vector $\tau$ generalized eigenvalue problem for Hamiltonian projection on subspace spanned by vectors $\exp{\tau}*e_J^I$ with $0\le |J|=|I|\le r$ is solved to give the ground state vector 
\begin{equation}
\Psi(\tau)=c_0\exp(\tau)+\sum\limits_{l=1}^r\sum\limits_{\genfrac{}{}{0pt}{}{J\subset R}{I\subset N-R}}^{(l)}c_J^I \exp(\tau)*e_J^I
\end{equation}
(ii) After dividing by  $c_0$ wave function is rewritten in the form
\begin{equation}
\Psi(\tau)=\exp(\tau+\Delta\tau)
\end{equation}
where
\begin{equation}
\Delta\tau=log\left (e_{\text{\O}}^{\text{\O}}+\sum\limits_{l=1}^r\sum\limits_{\genfrac{}{}{0pt}{}{J\subset R}{I\subset N-R}}^{(l)}c_J^Ie_J^I\right )
\end{equation}
(iii) If $\|\Delta\tau\|$ is still greater than some threshold value, we put $\tau \leftarrow\tau+\Delta\tau$ and return to step (i). 

At each step of this algorithm energy value  never rises. Note that at step (ii) we used the fact that exponent is an isomorphism of additive group of amplitudes onto multiplicative group ${\cal S}E^r_N$. For arbitrary polynomial parametrization the analogous algorithm is more complicated.

In the same simple manner it is possible to derive linearized equations for system evolution on arbitrary polynomial surface \textsl{in the vicinity of electronic energy stationary point}. To this end we should first get formula for electronic energy second derivatives assuming that the ground field is $\mathbb C$.  Realification of the amplitude space and calculation of the second energy derivatives with respect to real variables $x,y$ give
\begin{align}
D^2E_a(x,y)=\frac{2}{\|P_a(\tau)\|^2}\times\nonumber\qquad\quad\qquad\qquad\\
\left (\begin{array}{cc}
{\rm
Re}\left [\left ({\rm H}+\Delta\right )-E\left ({\rm G+S}\right )\right ]&-{\rm Im}\left [\left ({\rm H}+\Delta\right )-E\left ({\rm {G+S}}\right )\right ] \\
{\rm Im}\left [\left ({\rm H}-\Delta\right )-E\left (\rm {G-S}\right )\right ]&\ \  {\rm Re}\left [\left ({\rm H}-\Delta\right )-E\left (\rm {G-S}\right )\right ]
\end{array}\right )
\end{align}
where
\begin{subequations}
\begin{eqnarray}
{\rm H}_{JI,J'I'}=\langle P'_a(\tau)*e_J^I|H|P'_a(\tau)*e_{J'}^{I'}\rangle\\ 
{\rm G}_{JI,J'I'}=\langle P'_a(\tau)*e_J^I|P'_a(\tau)*e_{J'}^{I'}\rangle\\
\Delta_{JI,J'I'}=\langle P_a(\tau)|H|P''_a(\tau)*e_J^I*e_{J'}^{I'}\rangle\\
{\rm S}_{JI,J'I'}=\langle P_a(\tau)|P''_a(\tau)*e_J^I*e_{J'}^{I'}\rangle
\end{eqnarray}
\end{subequations}
Here ${\rm H}$ is a Hermitean matrix representing one of two identical blocks of projection of the electronic Hamiltonian on the realified tangent space to the surface ${\cal S}^r_N$ at a point $P_a(\tau)$, $\rm G$  is a Hermitean Gram (overlap) matrix of, in general non-orthogonal, basis $\{P'(\tau)*e_J^I\}$, $\Delta$ and $\rm S$ are {\it symmetric} matrices arising due to non-zero curvature of the aforementioned surface. 

Using standard technique based on the theory of Hamiltonian equations on symplectic manifolds (see, e.g. \cite{Rowe-1}-\cite{Rowe-4}),  it is easy to derive the following linearized evolution equation in variables $\tau,\bar{\tau}$ (Schrodinger form) 
\begin{eqnarray}
\left (\begin{array}{c}
\overset{.}\tau\\
\overset{.}{\bar \tau}
\end{array}\right )=
\frac{{\rm i}}{\|P_a(\tau_0)\|^2}\left (\begin{array}{cc}
\ {\rm H}-E_a(\tau_0){\rm G}&\bar\Delta-\bar S\\
-(\Delta +S)&\ -(\bar{{\rm H}}-E_a(\tau_0)\bar{{\rm G}})
\end{array}\right )
\left (\begin{array}{c}
\tau\\
\bar \tau
\end{array}\right )
\label{Schrod}
\end{eqnarray}
where $\tau_0$ is \textsl{ electronic energy stationary point}, matrices $\rm H$, $\rm G$, $\Delta$ and $\rm S$ are calculated at this point, and the amplitude vector $\tau$ in this equation is just the replacement vector from  $\tau_0$. In quantum chemistry such an approach to calculation of transition energies is called time-dependent (TD). Thus,  system (\ref{Schrod}) embraces all versions of TD methods based on polynomial parametrizations, including all versions of TDCC and TDQCI methods. 

In simple case of  parametrization $P_{CI}(\tau)=e_{\text{\O}}^{\text{\O}}+\tau$ Gram matrix reduces to the identity matrix, matrices $\Delta$ and $\rm S$ just vanish. And as could be expected, TDCI method, due to zero curvature of the surface ${\cal S}C^r_N$, gives nothing new in compare with the linear CI approach.

In concluding this section it is pertinent to give a very brief outline of possible generalization of variational problem under discussion. If MOs are to be varied, then the electronic energy domain may be elegantly described as a (locally trivial) vector bundle with Grassmann manifold over 1-electron sector of  Fock space as its base and formal amplitude vector space as a typical fibre. Such an approach embraces Hartree-Fock (HF) case (zero typical fiber) and Multi-Configurational Self-Consistent Field (MCSCF) methods along with already discussed ones, and all relevant TD methods. 

At present stage of development fully variational non-linear methods are considered as unfeasible. The reason is in problems with  Hamiltonian matrix element evaluation. Even if the current amplitude vector $\tau$ is of reasonably small length to be kept in fast memory, full CI expansion of $P_a(\tau)$ required for matrix element evaluation  may be of  huge size and level of  complexity of matrix element evaluation is close to that in the FCI method. One of the approaches which may help in avoiding this   bottleneck  is the following. It is possible to get complicated but closed analytic expression for vector $HP_a(\tau)$ \textsl{without preliminary FCI expansion of} $P_a(\tau)$. Arising on this way overlap-type matrix elements can be handled recurrently also without explicit CI expansions. However, this is only a hint for possible future developments.

\bigbreak \bigbreak
\newpage
{\Large \bf \noindent Polynomial Functions With  Quotient Algebras  As Their Range}
\bigbreak \bigbreak
For each fixed excitation level $r$ and each polynomial $P_a:W_N^r(R)\to { \rm A}_N^p(R)$ we can consider the function $P^r_a:W_N^r(R)\to { \rm A}_N^p(R)/{\rm I}_N^r(R)$ defined as
\begin{equation}
P^r_a : \tau\mapsto [P_a(\tau)]_r
\end{equation}
Using star multiplication in quotient algebra, we can write
\begin{equation}
P^r_a(\tau)=\sum\limits_{\mu=0}^ra_{\mu}[\tau^{\mu}]_r=\sum\limits_{\mu=0}^ra_{\mu}[\tau]_r^{\mu}
\end{equation}
Note that for calculation of polynomial value in quotient algebra only the first $r+1$ coefficients $a_{\mu}$ are  required. If $a_1\ne 0$ then  the inverse polynomial $[P_a^r]^{-1}$ exists and its coefficients $b_1,b_2,\ldots,b_r$ can be calculated using general Eq.(\ref{inv_coef}) ($b_0=0$ by definition).

Since Fock space is equipped with the inner product (see Eq.(\ref{innerp})), we can identify quotient space ${ \rm A}_N^p(R)/{\rm I}_N^r(R)$ with the orthogonal complement to ${\rm I}_N^r(R)$ which is just the subspace $\mathbb Ke_{\text{\O}}^{\text{\O}}(R)\oplus W_N^r(R)$.

Strictly speaking, for a fixed polynomial parametrization $P_a(\tau)$ the Schr{\"o}dinger equation   
\begin{equation}
HP_a(\tau)=EP_a(\tau)
\label{PolSchrod}
\end{equation}
has a solution if and only if 

(i) the exact eigenfunction of the electronic Hamiltonian $H$ has non-zero projection on the chosen reference subset $R$;

(ii) the excitation level is sufficiently high (ideally $r=\min\{p,n-p\}$).

In other words, the eigenvector should belong to the polynomial surface ${\cal S}^r_N$. In practice, however, relatively low excitations levels are used. This means that we can hardly hope to find amplitude vector, satisfying  equation (\ref{PolSchrod}).  But it is possible to consider quotient Schr{\"o}dinger equation in quotient algebra
\begin{equation}
[HP_a(\tau)]_r=E_a^r(\tau)[P_a(\tau)]_r
\label{q_PolSchrod}
\end{equation}
which can be recast in the following more habitual form
\begin{eqnarray}
\label{CCsys}
\langle e_J^I|HP_a(\tau)\rangle=E_a^r(\tau)\langle e_J^I|P_a(\tau)\rangle\qquad\\ 
(J\subset R, I\subset N-R, 0\le|J|=|I|\le r)\nonumber
\end{eqnarray}
The number of scalar equation in this system is $\dim [W_N^r(R)]+1$, that is exactly the number of unknowns (amplitude vector coordinates) plus 1. One additional equation corresponds to $J=I=\text{\O}$ and is used to evaluate 'energy' value for a fixed $\tau$:
\begin{equation}
E_a^r(\tau)=\frac{\langle e_{\text{\O}}^{\text{\O}}|HP_a(\tau)\rangle}{a_0}
\label{CCenergy}
\end{equation}
With such an approach 'energy' $E_a^r$ depends both on the parametrization (polynomial coefficients $a$) and excitation level ($r$), not being variational. It tends to the exact value when $r\rightarrow \min\{p,n-p\}$ \textsl{independently of} $a$. 

Iterative solution of  linearized system (\ref{CCsys}) is normally used in CC type calculations (see, e.g., \cite{Bartlett-3}). One can try as well to solve iteratively the initial vector equation (\ref{q_PolSchrod}) rewriting it in the form
\begin{equation}
[\tau]_r=[P_a^r]^{-1}\left (\frac{[HP_a(\tau)]_r}{E_a^r(\tau)}\right )=b_1\frac{[HP_a(\tau)]_r}{E_a^r(\tau)}+b_2\left [\frac{[HP_a(\tau)]_r}{E_a^r(\tau)}\right ]^2+\cdots
\end{equation}

In concluding this section it is pertinent to note that, when solving   the aforementioned equations, the main problem arising is the calculation of vector $[HP_a(\tau)]_r$ required at each iteration. To perform such a calculation CI expansion of the polynomial function $P_a(\tau)$ up to excitation level $r+2$ is required. Using star product,  we can instead derive  formulas for  direct action of creation-annihilation operator products on polynomial functions.

\bigbreak \bigbreak
{\Large \bf \noindent Computer Implementation Of Star Product In Quotient Algebras  }
\bigbreak \bigbreak

When using polynomial parametrization, one of the crucial points is the calculation of star product of two wave functions modulo $r$. Since almost all  calculations in quantum chemistry are performed in MO representation, from the very beginning we  separate $\alpha$ - and $\beta$ - indices and replace MSO index sets by pairs of MO index sets. In particular, MSO index set $N$ will be replaced by a pair $(M,M)$ of $\alpha$ - and $\beta$ - index sets, $|M|=m$, reference subset $R$ will become a pair $(R_{\alpha},R_{\beta})$, and basis elements of algebra ${ \rm A}_{(M,M)}^{(p_{\alpha},p_{\beta})}(R_{\alpha},R_{\beta})$ will be $e_{J_{\alpha},J_{\beta}}^{I_{\alpha},I_{\beta}}(R_{\alpha},R_{\beta})$. In what follows, unless otherwise indicated, it will be  supposed that the reference pair of index sets $(R_{\alpha},R_{\beta})$ is fixed, and    in all expressions dependence on $(R_{\alpha},R_{\beta})$ will be suppressed. 

Let us present classes of wave functions modulo $r$ to be $*$ -  multiplied in the standard form
\begin{subequations}
\begin{align}
[\Psi]_r={\rm I}^r + \sum\limits_{k=0}^r\sum\limits_{k_{\alpha}}\sum\limits_{\genfrac{}{}{0pt}{}{J_{\alpha}\subset R_{\alpha}}{I_{\alpha}\subset M-R_{\alpha}}}^{(k_{\alpha})}\sum\limits_{\genfrac{}{}{0pt}{}{J_{\beta}\subset R_{\beta}}{I_{\beta}\subset M-R_{\beta}}}^{(k_{\beta})}e_{J_{\alpha},J_{\beta}}^{I_{\alpha},I_{\beta}}
x_{J_{\alpha},J_{\beta}}^{I_{\alpha},I_{\beta}} \\
[\Phi]_r={\rm I}^r + \sum\limits_{k=0}^r\sum\limits_{k_{\alpha}}\sum\limits_{\genfrac{}{}{0pt}{}{J_{\alpha}\subset R_{\alpha}}{I_{\alpha}\subset M-R_{\alpha}}}^{(k_{\alpha})}\sum\limits_{\genfrac{}{}{0pt}{}{J_{\beta}\subset R_{\beta}}{I_{\beta}\subset M-R_{\beta}}}^{(k_{\beta})}e_{J_{\alpha},J_{\beta}}^{I_{\alpha},I_{\beta}}
y_{J_{\alpha},J_{\beta}}^{I_{\alpha},I_{\beta}} 
\end{align}
\end{subequations}
where $k_{\alpha}+k_{\beta}=k$ and $\max\{0,k-p_{\beta},k-m+p_{\beta}\}\le k_{\alpha}\le \min\{p_{\alpha},m-p_{\alpha},k\}$.
One of possible expressions for star product of two wave functions is
\begin{align}
[\Psi*\Phi]_r={\rm I}^r+\nonumber\qquad\qquad\qquad\qquad\qquad\qquad\qquad\\
\sum\limits_{k=0}^{r}\left \{\sum\limits_{k_{\alpha}}\sum\limits_{\genfrac{}{}{0pt}{}{J_{\alpha}\subset R_{\alpha}}{I_{\alpha}\subset M-R_{\alpha}}}^{(k_{\alpha})}\sum\limits_{\genfrac{}{}{0pt}{}{J_{\beta}\subset R_{\beta}}{I_{\beta}\subset M-R_{\beta}}}^{(k_{\beta})}e_{J_{\alpha},J_{\beta}}^{I_{\alpha},I_{\beta}}
\vphantom{\sum\limits_{k_{\alpha}=\max\{0,k-p_{\beta},k-n+p_{\beta}\}}^{\min\{k,p_{\alpha},n-p_{\alpha}\}}}\sum\limits_{k'=0}^k \sum\limits_{k'_{\alpha}}\sum\limits_{\genfrac{}{}{0pt}{}{J'_{\alpha}\subset J_{\alpha}}{I'_{\alpha}\subset I_{\alpha}}}^{(k'_{\alpha})}\sum\limits_{\genfrac{}{}{0pt}{}{J'_{\beta}\subset J_{\beta}}{I'_{\beta}\subset I_{\beta}}}^{(k'_{\beta})}(-1)^{\varepsilon}x_{J'_{\alpha},J'_{\beta}}^{I'_{\alpha},I'_{\beta}}y_{J_{\alpha}-J'_{\alpha},J_{\beta}-J'_{\beta}}^{I_{\alpha}-I'_{\alpha},I_{\beta}-I'_{\beta}}\right \}
\label{star_1}
\end{align}
where  $k'_{\alpha}+k'_{\beta}=k'$, $\max\{0,k'-k_{\beta}\}\le k'_{\alpha}\le \min\{k',k_{\alpha}\}$, and
\begin{equation}
\varepsilon=\sum\limits_{\sigma=\alpha,\beta}\left[|(J'_{\sigma}\cup I'_{\sigma})\cap \Delta_{(J_{\sigma}\cup I_{\sigma})}|+k'_{\sigma}\right ]
\end{equation}

Summation limits over $k_{\alpha}$ and $k'_{\alpha}$ require probably some comments. We have  inequalities 
\begin{eqnarray}
0\le k_{\alpha}\le p_{\alpha}\nonumber\\
0\le k_{\alpha}\le m-p_{\alpha}\nonumber\\
0\le k-k_{\alpha}\le p_{\beta}\nonumber\\
0\le k-k_{\alpha}\le m-p_{\beta}\nonumber
\end{eqnarray}
and, consequently, $k_{\alpha}\ge \max\{0,k-p_{\beta},k-m+p_{\beta}\}$, $k_{\alpha}\le \min\{k, p_{\alpha}, m-p_{\alpha}\}$. As for $k'_{\alpha}$, we have 
\begin{eqnarray*}
0\le k'_{\alpha}\le k_{\alpha}\\
0\le k'-k'_{\alpha}\le k_{\beta}
\end{eqnarray*}
which means that $\max\{0,k'-k_{\beta}\}\le k'_{\alpha}\le \min\{k',k_{\alpha}\}$. Simple part of star product in Eq.(\ref{star_1}) corresponds to $k'=0,k$ and is $x_{\text{\O},\text{\O}}^{\text{\O},\text{\O}}\Phi+\Psi y_{\text{\O},\text{\O}}^{\text{\O},\text{\O}}$. 

The second version of formula for star product is 
\begin{align}
[\Psi*\Phi]_r={\rm I}^r+
\sum\limits_{k=0}^{r}\sum\limits_{k_{\alpha}}\sum\limits_{\genfrac{}{}{0pt}{}{J_{\alpha}\subset R_{\alpha}}{I_{\alpha}\subset M-R_{\alpha}}}^{(k_{\alpha})}\sum\limits_{\genfrac{}{}{0pt}{}{J_{\beta}\subset R_{\beta}}{I_{\beta}\subset M-R_{\beta}}}^{(k_{\beta})}x_{J_{\alpha},J_{\beta}}^{I_{\alpha},I_{\beta}}\times\nonumber\\
\left [ 
\vphantom{\sum\limits_{\genfrac{}{}{0pt}{}{J'_{\beta}\subset R_{\beta}-J_{\beta}}{I'_{\beta}\subset N-R_{\beta}-I_{\beta}}}^{(k'_{\beta})}e_{J_{\alpha}\cup J'_{\alpha},J_{\beta}\cup J'_{\beta}}^{I_{\alpha}\cup I'_{\alpha},I_{\beta}\cup I'_{\beta}}y_{J'_{\alpha},J'_{\beta}}^{I'_{\alpha},I'_{\beta}}(-1)^{\sum\limits_{\sigma=\alpha,\beta}|(J'_{\sigma}\cup I'_{\sigma})\cap \Delta_{(J_{\sigma}\cup I_{\sigma})}|}}
\sum\limits_{k'=0}^{r-k}
\sum\limits_{k'_{\alpha}}\sum\limits_{\genfrac{}{}{0pt}{}{J'_{\alpha}\subset R_{\alpha}-J_{\alpha}}{I'_{\alpha}\subset M-R_{\alpha}-I_{\alpha}}}^{(k'_{\alpha})}\sum\limits_{\genfrac{}{}{0pt}{}{J'_{\beta}\subset R_{\beta}-J_{\beta}}{I'_{\beta}\subset M-R_{\beta}-I_{\beta}}}^{(k'_{\beta})}e_{J_{\alpha}\cup J'_{\alpha},J_{\beta}\cup J'_{\beta}}^{I_{\alpha}\cup I'_{\alpha},I_{\beta}\cup I'_{\beta}}(-1)^{\varepsilon}y_{J'_{\alpha},J'_{\beta}}^{I'_{\alpha},I'_{\beta}}\right ]
\end{align}
\begin{equation}
\varepsilon=\sum\limits_{\sigma=\alpha,\beta}|(J'_{\sigma}\cup I'_{\sigma})\cap \Delta_{(J_{\sigma}\cup I_{\sigma})}|
\end{equation}
In the above equations $r$ - maximal excitation level used. For full (not quotient)  product $r=\min\{p,2m-p\}$. Note that simple contributions in the last expression correspond to $k=0$ and $k'=0$. 

At present we are using the second expression for the star product modulo $r$ because in our realization the corresponding algorithm works faster than that based on the first expression. 

Any computer implementation of start product requires  algorithm for generation of subsets with fixed number of elements contained in  a given index set. For this there exist variety of  methods which differ in subset orderings: lexical, Gray, revolving door, etc. Apart from generation itself, very important also is availability of the corresponding ranking and unranking functions. 

After series of experiments we came to the conclusion that the revolving door algorithm is the best  for our purpose. This algorithm and accompanying ranking and unranking functions are described in \cite{Skiena}. The corresponding codes are free and can be found at \textsl{ The Stony Brook Algorithm Repository} \cite{algorithm}. 

For our current purpose unranking function is of no use, but ranking one is compulsory. Let us denote by  ${\rm rank}_RJ$  rank of $J$ as a subset of $R$ in revolving door ordering, and by ${\rm rank}_{(R_{\alpha},R_{\beta})}(J_{\alpha},J_{\beta})$ rank of a pair of subsets calculated as
\begin{equation}
{\rm rank}_{(R_{\alpha},R_{\beta})}(J_{\alpha},J_{\beta})=\left ({\rm rank}_{R_{\alpha}}J_{\alpha}-1\right )\binom{p_{\alpha}}{k_{\alpha}}+{\rm rank}_{R_{\beta}}J_{\beta}
\end{equation}
Then for given $k$, $k_{\alpha}$, $J_{\alpha}$, $J_{\beta}$, $I_{\alpha}$, and $I_{\beta}$ the position of the corresponding element in a row vector representing wave function (as well as  its amplitude) is determined with the aid of the following mapping: 
\begin{align}
(k, k_{\alpha}, J_{\alpha}, J_{\beta}, I_{\alpha}, I_{\beta})\mapsto {\rm offset}(k,k_{\alpha})+\nonumber\qquad\qquad\qquad\\
\left ({\rm rank}_{(R_{\alpha},R_{\beta})}(J_{\alpha},J_{\beta})-1\right )\binom{m-p_{\alpha}}{k_{\alpha}}\binom{m-p_{\beta}}{k_{\beta}}+{\rm rank}_{(M-R_{\alpha},M-R_{\beta})}(I_{\alpha},I_{\beta})
\label{ordering}
\end{align}
and where the array of offset indices (filled only once at the beginning of calculation) is 
\begin{align}
{\rm offset}(k,k_{\alpha})=\sum\limits_{k'=1}^{k-1}\sum\limits_{k'_{\alpha}=\max\{0,k'-p_{\beta},k'-m+p_{\beta}\}}^{\min\{k',p_{\alpha},m-p_{\alpha}\}}\binom{p_{\alpha}}{k'_{\alpha}}\binom{m-p_{\alpha}}{k'_{\alpha}}\binom{p_{\beta}}{k'_{\beta}}\binom{m-p_{\beta}}{k'_{\beta}}+\nonumber\\
\sum\limits_{k'_{\alpha}=\max\{0,k-p_{\beta},k-m+p_{\beta}\}}^{k_{\alpha}-1}\binom{p_{\alpha}}{k'_{\alpha}}\binom{m-p_{\alpha}}{k'_{\alpha}}\binom{p_{\beta}}{k'_{\beta}}\binom{m-p_{\beta}}{k'_{\beta}}\qquad\qquad
\end{align}
Computer code was written in FORTRAN 95 and was heavily tested. Example of time characteristics of current 
computer implementation of star product modulo $r$ are given in Table \ref{tab:timing}.
\begin{table}[h]
	\centering
			\begin{tabular}{crr}
\rule{0pt}{1pt}\\
\hline
\rule{0pt}{1pt}\\
\multicolumn{1}{c}{Excitation}     &\multicolumn{1}{c}{Dimension of}  &\multicolumn{1}{c}{Time} \\
\multicolumn{1}{c}{level}          &\multicolumn{1}{c}{the amplitude}     &\multicolumn{1}{c}{(sec)} \\
 \multicolumn{1}{c}{$r$}             &\multicolumn{1}{c}{space}         &      \\
\rule{0pt}{1pt}\\
\hline
\rule{0pt}{15pt}
3 &18818     &0.02\\  
4 &98693     &0.35\\  
5 &294965    &2.75\\  
6 &558809    &13.04\\  
7 &755081    &38.49\\  
8 &834956    &74.62\\
9 &   851956 & 103.10  \\
 10&  853702 & 115.27 \\
 11&   853774 & 117.95  \\
 12&  853775  & 118.23 \\
\hline
		\end{tabular}
	\caption{Timing of star product modulo $r$ calculations (Intel Fortran, CPU Q9550@2.83GHz, sequential mode).}
	\label{tab:timing}
\end{table}

\bigbreak \bigbreak  
{\Large \bf \noindent Comparison of Different Polynomial }

{\Large \bf \noindent Parametrizations On Concrete Examples }
\bigbreak \bigbreak

Let us suppose that for a molecule under consideration orthonormal MO set is generated at the HF level of theory, active space is chosen, and FCI calculation of few low-lying electronic states  is performed  in this active space. Let  $\Psi_{FCI}$ be one of the calculated wave functions. Our nearest aim is to compare this exact wave function with its approximation by different polynomial functions for different excitation levels. To this end we first use  GAMESS US  program \cite{GAMESS} (ALDET route) to generate coefficients of FCI expansion over determinant basis set. Then maximal by absolute value FCI coefficient is found and the corresponding determinant is taken as the reference one. Next step is to reorder FCI coefficients in the ordering defined by the mapping (\ref{ordering}) (in GAMESS lexical ordering is accepted). Then FCI wave function is divided by the coefficient $c_0=c_{\text{\O},\text{\O}}^{\text{\O},\text{\O}}$ and for a selected polynomial parametrization \textsl{the exact} amplitude vector is determined:
\begin{equation}
\tau_{a}^{exact}=P^{-1}_a(e_{\text{\O},\text{\O}}^{\text{\O},\text{\O}}+x)
\end{equation}
where $x=\frac{1}{c_0}\Psi_{FCI}-e_{\text{\O},\text{\O}}^{\text{\O},\text{\O}}$. 'Exact` means that $P_a(\tau_{a}^{exact})=\frac{1}{c_0}\Psi_{FCI}$.  This amplitude vector depends on the parametrization used, carries therefore subscript $a$, and  involves, in general,   all possible excitations from the reference state.  

There are two questions arising. 

(i) Suppose that the exact amplitude vector is truncated  to the level $r$, and full expansion of the wave function $P_a(\tau)$  is calculated. The question is: At what excitation level polynomial of exact but truncated amplitude  becomes reasonably close to the FCI function? 

(ii) Suppose that truncated amplitude vector is optimized to make $\|P_a(\tau)-\Psi_{FCI}\|$ as small as possible. The question is: In what cases optimization may  essentially improve polynomial approximation of $\Psi_{FCI}$? 

Of quite a large number of performed  atomic and molecular calculations in different bases and with different active spaces we give here the results  for only three molecular systems: (1)  calculation of  hypothetic molecule ${\rm H_2C-NH-CHF}$ ground state which is strongly correlated; (2) calculation of ${\rm N_2O}$ molecule ground state; (3) calculation of  the first excited (triplet) state of  boric acid molecule (its ground state is of no interest for us since it is practically non-correlated). 

\begin{table}[ht]
	\centering
\begin{tabular}{ccccc}
\rule{0pt}{1pt}\\
\hline
\rule{0pt}{1pt}\\
Polynomial&Excitation&Dimension of the&$\|P_a(\tau)-\Psi_{FCI}\|$&$\|P_a(\tau^*)-\Psi_{FCI}\|$\\ 
$P_a(\tau)$&level& amplitude space&&\\
\rule{0pt}{1pt}\\
\hline
\rule{0pt}{15pt}
$\exp(\tau)$
&   1&       64 &0.420632E+00 &0.417226E+00 \\
&   2&     1424 &0.153643E-01 &0.144873E-01 \\
&   3&    12624 &0.207110E-02 &0.191069E-02 \\
&   4&    55324 &0.123119E-03 &0.113253E-03 \\
&   5&   135068 &0.114390E-04 &0.114390E-04 \\
&   6&   208764 &0.375204E-06 &0.375204E-06 \\
\hline
$Q(\tau)$
&   1&        64 &0.438430E+00 &0.425482E+00 \\
&   2&      1424 &0.550005E-01 &0.405679E-01 \\
&   3&     12624 &0.270095E-01 &0.194293E-01 \\
&   4&     55324 &0.183023E-02 &0.135976E-02 \\
&   5&    135068 &0.746502E-03 &0.532138E-03 \\
&   6&    208764 &0.118016E-04 &0.888355E-05 \\
\hline
$P_{CI}(\tau)$
&   1&        64 &0.406908E+00  &  \\
&   2&      1424 &0.396744E-01  &  \\
&   3&     12624 &0.249536E-01  &  \\
&   4&     55324 &0.112012E-02  &  \\
&   5&    135068 &0.709024E-03  &  \\
&   6&    208764 &0.872110E-05  &  \\
\hline
$q_{\frac{1}{2}}(\tau)$
&   1&       64 &0.420542E+00 &0.417171E+00 \\
&   2&     1424 &0.163069E-01 &0.151996E-01 \\
&   3&    12624 &0.390887E-02 &0.359143E-02 \\
&   4&    55324 &0.157044E-02 &0.139820E-02 \\
&   5&   135068 &0.917063E-03 &0.815560E-03 \\
&   6&   208764 &0.438127E-04 &0.389069E-04 \\
\hline
		\end{tabular}
	\caption{Comparison of different parametrizations: Ground state of molecule $\rm H_2C-NH-CHF$.}
	\label{tab:ylide}
\end{table}

\begin{table}[ht]
	\centering
\begin{tabular}{ccccc}
\rule{0pt}{1pt}\\
\hline
\rule{0pt}{1pt}\\
Polynomial&Excitation&Dimension of the&$\|P_a(\tau)-\Psi_{FCI}\|$&$\|P_a(\tau^*)-\Psi_{FCI}\|$\\ 
$P_a(\tau)$&level& amplitude space&&\\
\rule{0pt}{1pt}\\
\hline
\rule{0pt}{15pt}
$\exp(\tau)$
&   1&        72 &0.415883D+00 &0.400344D+00 \\
&   2&      1818 &0.110642D+00 &0.106583D+00 \\
&   3&     18818 &0.415729D-01 &0.392551D-01 \\
&   4&     98693 &0.938115D-02 &0.887311D-02 \\
&   5&    294965 &0.803279D-03 &0.744422D-03 \\
&   6&    558809 &0.450764D-04 &0.412938D-04 \\
\hline
$Q(\tau)$
&   1&        72 &0.628926D+00 &0.4495243D+00 \\
&   2&      1818 &0.347899D+00 &0.1994256D+00 \\
&   3&     18818 &0.164148D+00 &0.9625347D-01 \\
&   4&     98693 &0.599690D-01 &0.3734982D-01 \\
&   5&    294965 &0.168746D-01 &0.1129635D-01 \\
&   6&    558809 &0.191710D-02 & 0.1073262D-02\\
\hline
$P_{CI}(\tau)$
&   1&        72 &0.369100D+00  &  \\
&   2&      1818 &0.171212D+00  &  \\
&   3&     18818 &0.727427D-01  &  \\
&   4&     98693 &0.211776D-01  &  \\
&   5&    294965 &0.435432D-02  &  \\
&   6&    558809 &0.556137D-03  &  \\
\hline
$q_{\frac{1}{2}}(\tau)$
&   1&        72 &0.405992E+00 &0.394752E+00 \\
&   2&      1818 &0.119679E+00 &0.110201E+00 \\
&   3&     18818 &0.822454E-01 &0.579319E-01 \\
&   4&     98693 &0.685212E-01 &0.418136E-01 \\
&   5&    294965 &0.446974E-01 &0.281541E-01 \\
&   6&    558809 &0.135399E-01 &0.771915E-02 \\
\hline
		\end{tabular}
	\caption{Comparison of different parametrizations: Ground state of molecule $\rm N_2O$.}
	\label{tab:n2o}
\end{table}

\begin{figure}[ht]
	\centering
		\includegraphics{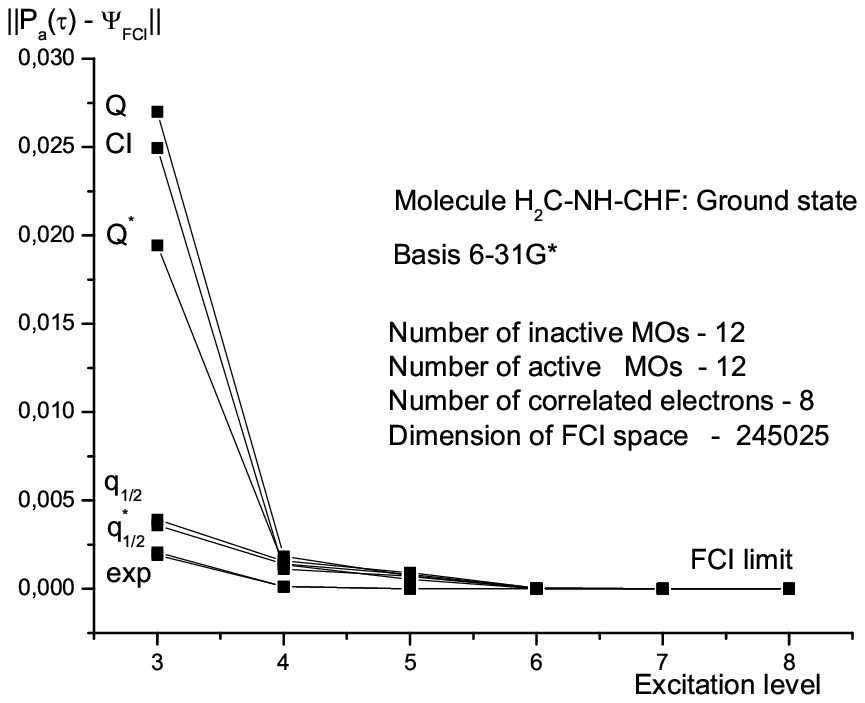}
	\caption{Comparision of different polynomial approximations on example of the ground state of molecule
${\rm H_2C-NH-CHF}$. Curves corresponding to optimized amplitudes carry additional label *.}
	\label{fig:ehf}
\end{figure}
\begin{figure}[ht]
	\centering
		\includegraphics{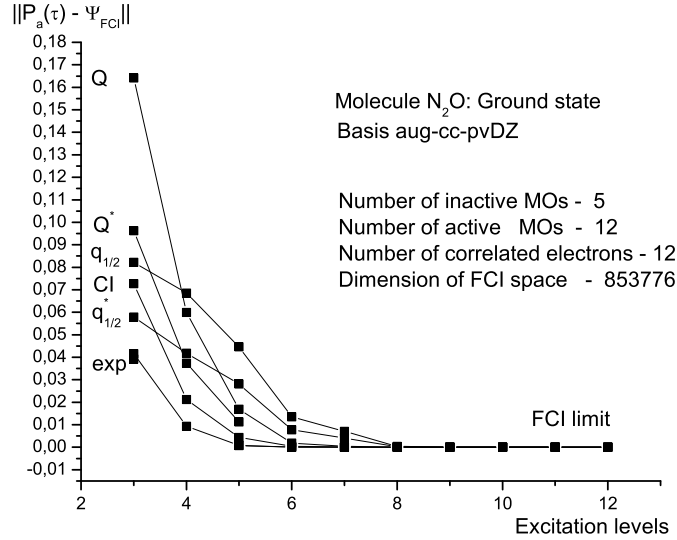}
	\caption{Comparision of different polynomial approximations on example of the ground state of molecule
${\rm N_2O}$. Curves corresponding to optimized amplitudes carry additional label *.}
	\label{fig:n2o}
\end{figure}

\begin{figure}[ht]
	\centering
		\includegraphics{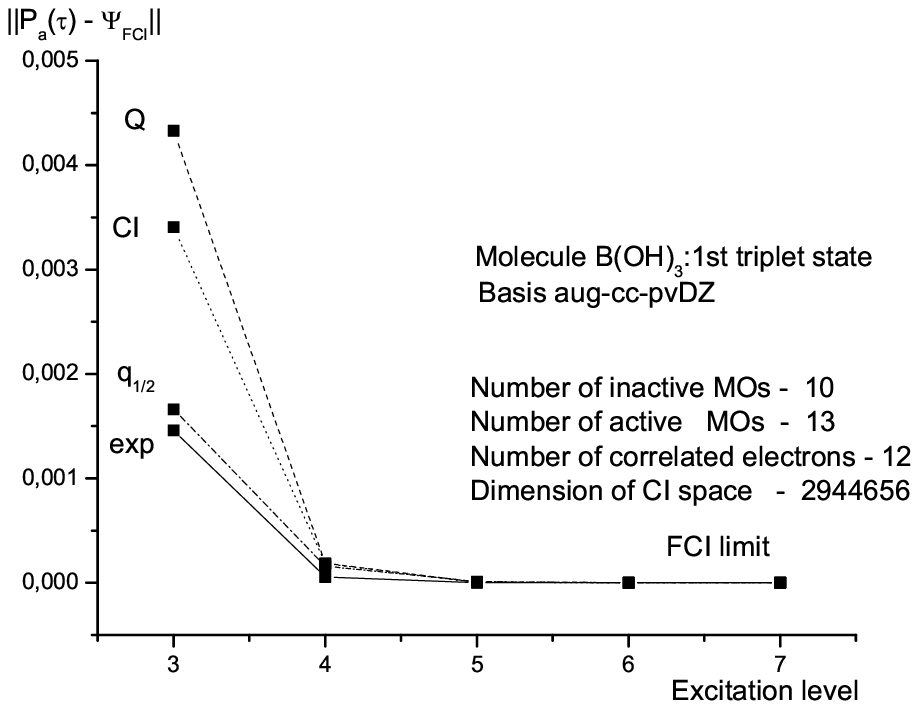}
	\caption{Comparision of different polynomial approximations on example of the 1st triplet state of boric acid molecule.}
	\label{fig:boric}
\end{figure}

Comparision of quality for different polynomial appriximations of FCI functions for the aforementioned molecules are depicted in Figs.(\ref{fig:ehf})-(\ref{fig:boric}) and presented in Tables (\ref{tab:ylide})-(\ref{tab:n2o}).
 From Figures and Tables it is seen that {\it all} polynomial parametrizations give reasonably good approximations of FCI functions  starting from the excitation level 6. For lower excitation levels exponential parametrization is beyond comparision giving the best results. Behaviour of polynomial approximations in the course of amplitude optimization is of special interest. Such optimization may essentially improve wave function approximation for, say,  resolvent  ($Q$ polynomial) or quadratic ($q_{\frac{1}{2}}$ polynomial) parametrizations. In the CI case optimal truncated amplitude vector just coincides with the truncated part of the CI expansion divided by $c_0$. And   it seems that the exponential parametrization inherits this property of truncated CI expansions. Namely, it possesses the following  for the time being empirical but important property:

\textsl{For exponential parametrization optimal amplitude vector involving excitations up to level $r$ is a very good approximation to the exact amplitude vector truncated to the same excitation level.  }

It is to be noted that quadratic parametrization gives, as a rule, approximations of wave functions close in their quality to the exponential parametrization. However, in a number of cases behaviour of such simple parametrizations may be  strange and unpredictable (see Fig.(\ref{fig:n2o})). It is therefore necessary to keep in mind that quality of simple parametrizations (quadratic, qubic, etc.) may be system dependent.

And in concluding this section it is pertinent to mention that all parametrizations with excitation levels 1 and 2 in the amplitude vector  give poor approximations of wave functions as seen from Tables (\ref{tab:ylide})-(\ref{tab:n2o}).

\bigbreak \bigbreak
{\Large \bf \noindent Wave Function of  Non-Interacting Subsystems }
\bigbreak \bigbreak

We confine ourselves to the case of two non-interacting  subsystems.

Let us assume that the first $n_1$ vectors $e_i$ span MSO space of subsystem I, and the remainder vectors $e_i$ span MSO space of subsystem II. The 'one-electron' sector of formal Fock space is a direct sum 
\begin{equation}
{\mathfrak F}^1_{N}={\mathfrak F}^1_{X_1}\oplus {\mathfrak F}^1_{X_2}
\end{equation}
where $X_1=\{1,2,\ldots,n_1\}$ and $X_2=\{n_1+1,n_1+2,\ldots,n\}$.

Let $\Psi_I\in {\mathfrak F}^{p_1}_{X_1}$ and $\Psi_{II}\in {\mathfrak F}^{p_2}_{X_2}$. Wave function describing system of two non-interacting subsystems is just the Grassmann product  $\Psi_{I+II}=\Psi_I\wedge\Psi_{II}\in {\mathfrak F}^{p_1+p_2}_N$. Is it possible to present such a wave function as a star product of $\Psi_I$ and $\Psi_{II}$?  The answer is positive. Indeed, let us select reference subsets $R_1\subset X_1$ and $R_2\subset X_2$ for states $\Psi_I$ and $\Psi_{II}$, respectively, and expand these states in a standard manner
\begin{subequations}
\begin{eqnarray}
\Psi_I(R_1)=\sum\limits_{l_1=0}^{\min\{p_1,n_1-p_1\}}\sum\limits_{\genfrac{}{}{0pt}{}{J_1\subset R_1}{I_1\subset X_1-R_1}}c_{J_1}^{I_1}e_{J_1}^{I_1}(R_1)\\
\Psi_{II}(R_2)=\sum\limits_{l_2=0}^{\min\{p_2,n_2-p_2\}}\sum\limits_{\genfrac{}{}{0pt}{}{J_2\subset R_2}{I_2\subset X_2-R_2}}c_{J_2}^{I_2}e_{J_2}^{I_2}(R_2)
\end{eqnarray}
\end{subequations}
Since, by definition, $e_{J_i}^{I_i}(R_i)=e_{\text{\O}}^{R_i-J_i\cup I_i}(\text{\O})$, we can write
\begin{equation}
e_{\text{\O}}^{R_1-J_1\cup I_1}(\text{\O})\wedge e_{\text{\O}}^{R_2-J_2\cup I_2}(\text{\O})=e_{\text{\O}}^{R-J\cup I}(\text{\O})
\end{equation}
where $R=R_1\cup R_2$, $J=J_1\cup J_2$, and $I=I_1\cup I_2$. The sign on the right-hand side of the last equality is always plus due to the chosen ordering of MSO indices. But due to the same ordering we also have 
\begin{equation}
e_{\text{\O}}^{R-J\cup I}(\text{\O})=e_{J_1}^{I_1}(R)*e_{J_2}^{I_2}(R)
\end{equation}
Immediate consequence of this relation and multiplication law (\ref{multtab}) is the following  equality
\begin{equation}
\Psi_I(R_1)\wedge \Psi_{II}(R_2)=\Psi_I(R)*\Psi_{II}(R)
\label{wedge_star}
\end{equation}
In fact, in the last equality we used implicitely the embedding  of $q$ - electron algebra ${ \rm A}_{X}^{q}(S)$ into $p$ - electron algebra ${ \rm A}_N^p(R)$  ($q\le p)$.

With words the equality (\ref{wedge_star})  can be explained in the following way. It is necessary to take wave functions describing non-interacting systems and expand them using their reference subsets and their MSO index sets. Then  each of these  functions     should be embedded into $p$ - electron sector of the Fock space where united reference set $R_1\cup R_2$ is chosen, and then star product should be calculated.

\bigbreak \bigbreak
{\Large \bf \noindent Conclusion }
\bigbreak \bigbreak
      
This work is completing  the construction of general algebraic theory of non-linear methods of quantum chemistry which was started in our previous publications. Efficient computer implementation of star product for the most general case made it possible to compare different polynomial approximations of many-electron wave functions for different excitations levels. Further possible generalizations embracing theories with MO optimization require essentially more complicated mathematical tools of modern differential geometry, in particular, theory of vector bundles. On this route it is possible to construct theory including practically all existing wave function based  variational methods of quantum chemistry. 

But in the frameworks of described in this work purely algebraic approach there still exits one unsolved fundamental problem. This problems concerns the question of consistency between star product and the standard Hermitean norm on the Fock space. In more rigourous terms, it is desirable to ascertain that the inequality $\|\Psi*\Phi\|\le \|\Psi\|\|\Phi\|$ holds true for arbitrary $p$ - electron wave functions when the number of MSOs is sufficiently large. If this statement is correct, then practically all the results  described in this work can be easily reformulated for infinite-dimensional Fock spaces and then the star product will become the property of quantum mechanics but not only of its finite dimensional model which is at present quantum chemistry.
\newpage
 \bigbreak

\end{document}